\documentclass[conference]{IEEEtran}

\usepackage{cite}
\usepackage{amsmath,amssymb,amsfonts}
\usepackage{graphicx}
\usepackage{textcomp}
\usepackage{xcolor}
\usepackage{booktabs}
\usepackage{multirow}
\usepackage{array}
\usepackage{url}
\IEEEoverridecommandlockouts

\begin{document}

\title{Embodied Multimedia: A Tutorial}

\author{
Yang Liu$^{1,\dagger}$, Wei Zuo$^{2}$, Guanwei Zhao$^{1}$, Juncen Guo$^{2,*}$, Jiangchuan Liu$^{3}$, Abdulmotaleb Saddik$^{4}$, Liang Song$^{2}$ \\
$^{1}$Tongji University \quad $^{2}$Fudan University \quad $^{3}$Simon Fraser University \quad $^{4}$University of Ottawa   \\
{\small \{yangliu25,2451329\}@tongji.edu.cn, \{wzuo25,guojc23\}@m.fudan.edu.cn, jcliu@sfu.ca, elsaddik@uottawa.ca, songl@fudan.edu.cn}
\thanks{$^\dagger$This project is lead by the Embodied Perception \& Intelligence Computing Lab, Tongji University (https://tongji-epic.github.io/). $^*$Corresponding author.}
}

\maketitle

\begin{abstract}
Traditional multimedia technology has been built around optimizing content delivery for human observers, from perceptually driven compression standards to human-centric quality metrics. With the rapid rise of embodied intelligence, autonomous agents must perceive, reason, and act within the physical world in real time, exposing fundamental mismatches between conventional multimedia infrastructure and the demands of embodied tasks. In this regard, this tutorial paper formally introduces Embodied Multimedia as a cross-disciplinary research paradigm that treats multimodal data as the perceptual and communicative substrate spanning the full perception-decision-action loop. To be specific, we present a four-layer unified architecture comprising Data, Communication, Cognitive, and Evaluation layers, and provide a structured review of key enabling technologies within each layer. Furthermore, we identify five frontier application directions where Embodied Multimedia is positioned to serve a foundational role: multimedia communication, physical intelligence, embodied anomaly perception, the metaverse and interactive multimedia, and AI-driven art creation. Open technical challenges and future research directions are discussed to guide the community in this emerging field.
\end{abstract}


\section{Introduction}

Multimedia computing has served as the foundational infrastructure of the modern information age for several decades. By integrating diverse data modalities (e.g., text, images, audio, video, and 3D content), it has established mature technical pipelines spanning acquisition, compression, transmission, retrieval, generation, and quality assessment. These pipelines share a unifying design principle: optimizing the sensory experience for \textit{human observers}. Compression standards discard information that falls below human perceptual thresholds. Evaluation metrics measure signal fidelity relative to human visual sensitivity. Content generation systems, from early GANs~\cite{goodfellow2020gans} to modern diffusion models~\cite{sohldickstein2015diffusion} and pre-trained Transformers, aim primarily to produce aesthetically coherent outputs for human consumption.

The rise of Embodied Intelligence (EI) is fundamentally altering the landscape of multimedia research~\cite{sun2024embodied}. Unlike conventional AI that operates over pre-collected static datasets in cyberspace, embodied agents are equipped with physical bodies and must continuously perceive, reason, and act within dynamic real-world environments. The perception-decision-action loop governing such agents, spanning robotic manipulators, autonomous vehicles, humanoid robots, and virtual agents in extended reality, places entirely different demands on data: rather than delivering compelling sensory experiences, data must support reliable machine perception, enable physical reasoning, and underpin safe and precise manipulation.

\begin{figure}[t]
\centering
\includegraphics[width=0.8\columnwidth]{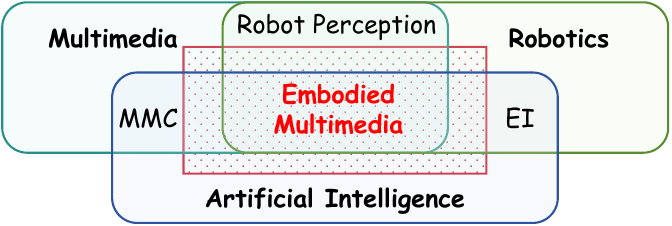}
\caption{Relationship among AI, Multimedia Computing (MMC), Embodied Intelligence, and the proposed EMM paradigm. EMM occupies the intersection of multimedia infrastructure and embodied agent requirements, providing the task-driven data substrate spanning the perception-decision-action loop.}
\label{fig:relations}
\vspace{-15pt}
\end{figure}

A critical distinction, illustrated in Fig.~\ref{fig:relations}, separates \textit{Embodied Intelligence (EI)} from \textit{Embodied Multimedia (EMM)}. EI defines the cognitive and behavioral architecture of agents, specifying how they plan, reason, and act. EMM, in contrast, addresses the multimodal data and communication infrastructure that makes EI operational: how perceptual data is acquired, compressed, transmitted, retrieved, generated, and evaluated in a manner aligned with agent task objectives rather than human sensory preferences. Just as traditional multimedia provides the data substrate for human-centered applications, EMM serves as the perceptual and communicative substrate for embodied agents, spanning the full perception-decision-action loop. Without an appropriate multimedia infrastructure, EI systems are fundamentally constrained: compression sacrifices task-relevant features, transmission latency breaks closed-loop control, and evaluation criteria do not reflect physical task performance. 

In summary, the contributions of this tutorial paper include: (1) We formally define the EMM paradigm and characterize its distinction from both traditional multimedia and embodied intelligence. (2) We propose a four-layer unified architecture comprising Data, Communication, Cognitive, and Evaluation layers. (3) We survey key enabling technologies across all four layers. (4) We identify five frontier application directions where EMM is positioned to play a foundational role. We believe this paper provides both a conceptual foundation and a practical roadmap for the research community at the intersection of multimedia computing and embodied intelligence.

\section{Limitations Analysis and motivation statement}
\label{sec:limitations}

The design assumptions of conventional multimedia systems conflict with the operational requirements of embodied agents across the following five key dimensions. Table~\ref{tab:comparison} summarizes the contrasts between traditional and Embodied Multimedia.

\textit{(1) Data acquisition and processing.} Traditional systems passively capture data from fixed or predetermined viewpoints, prioritizing global scene coverage over task-relevant semantic regions. The static paradigm prevents agents from resolving occlusions, directing sensor attention, or exploring unknown semantic areas. Standard lossy codecs are designed around human visual system models and remove high-frequency components imperceptible to humans but critical for machine perception, including the fine-grained edge and texture detail that underpins manipulation and spatial reasoning~\cite{li2025machine}.

\textit{(2) Communication.} Shannon-theoretic frameworks strive for reliable bit-level delivery of raw data streams through unidirectional, centralized pipelines. Embodied control requires multimodal streams synchronized at millisecond timescales, which such architectures cannot support. Centralized cloud processing further aggravates this latency problem, making closed-loop control impractical under realistic conditions.

\textit{(3) Retrieval.} Content-based retrieval builds unified feature spaces for similarity-driven matching, which is effective for explicit human queries but insufficient when embodied agents must reason over ambiguous instructions, decompose long-horizon tasks, and update their knowledge state during execution. Traditional systems lack the closed-loop mechanisms for autonomous task analysis and feedback-driven refinement.

\textit{(4) Generative modeling.} Conventional multimedia generation is designed for digital aesthetics. Models trained to produce realistic images and videos lack grounding in physical laws, material properties, and contact geometry. Embodied agents require generative systems capable of producing actionable outputs, such as physically plausible future environmental states for planning, or executable control sequences.

\textit{(5) Quality evaluation.} Metrics such as PSNR and SSIM quantify signal fidelity relative to human perceptual sensitivity but do not reflect the decision-making utility of data for machine tasks. Measures for physical interaction safety, spatial reasoning accuracy, and long-horizon task completion are largely absent from conventional evaluation frameworks.

\begin{table}[t]
\centering
\caption{Comparison of Traditional and Embodied Multimedia}
\label{tab:comparison}
\setlength{\tabcolsep}{3pt}
\renewcommand{\arraystretch}{1.25}
\begin{tabular}{p{1.8cm}p{3.0cm}p{3.2cm}}
\toprule
\textbf{Dimension} & \textbf{Traditional Multimedia} & \textbf{Embodied Multimedia} \\
\midrule
Target        & Human observers           & Embodied agents              \\
Core goal     & Perceptual fidelity       & Task execution               \\
Acquisition   & Passive, fixed viewpoints & Active, adaptive sensing     \\
Processing    & Human-perceptual coding   & Semantic-preserving coding   \\
Communication & Bit-level, centralized    & Task-oriented, edge-cloud    \\
Retrieval     & Static feature matching   & Agent-based dynamic loop     \\
Generation    & Aesthetic digital content & Actionable physical outputs  \\
Evaluation    & PSNR, SSIM, LPIPS           & Task success, safety metrics \\
\bottomrule
\end{tabular}
\vspace{-20pt}
\end{table}

\section{Definition and Architecture of EMM}
\label{sec:emm}

\subsection{Definition}

Embodied Multimedia is a multimedia computing paradigm designed for embodied intelligent agents. EMM treats multimodal data not as content for passive human consumption, but as the perceptual and communicative substrate spanning the full perception-decision-action loop. The paradigm covers data acquisition and processing, semantic transmission, cognitive decision-making, and task-aligned evaluation. Each component is oriented toward enabling agents to perceive physical environments reliably, form physically grounded plans, and execute physical actions effectively. Embodied Multimedia extends traditional multimedia beyond sensory delivery to encompass the dynamic interplay between data, communication, and physical execution in open-world environments.

\subsection{Unified Architecture}

\begin{figure}[t]
\centering
\includegraphics[width=0.99\columnwidth]{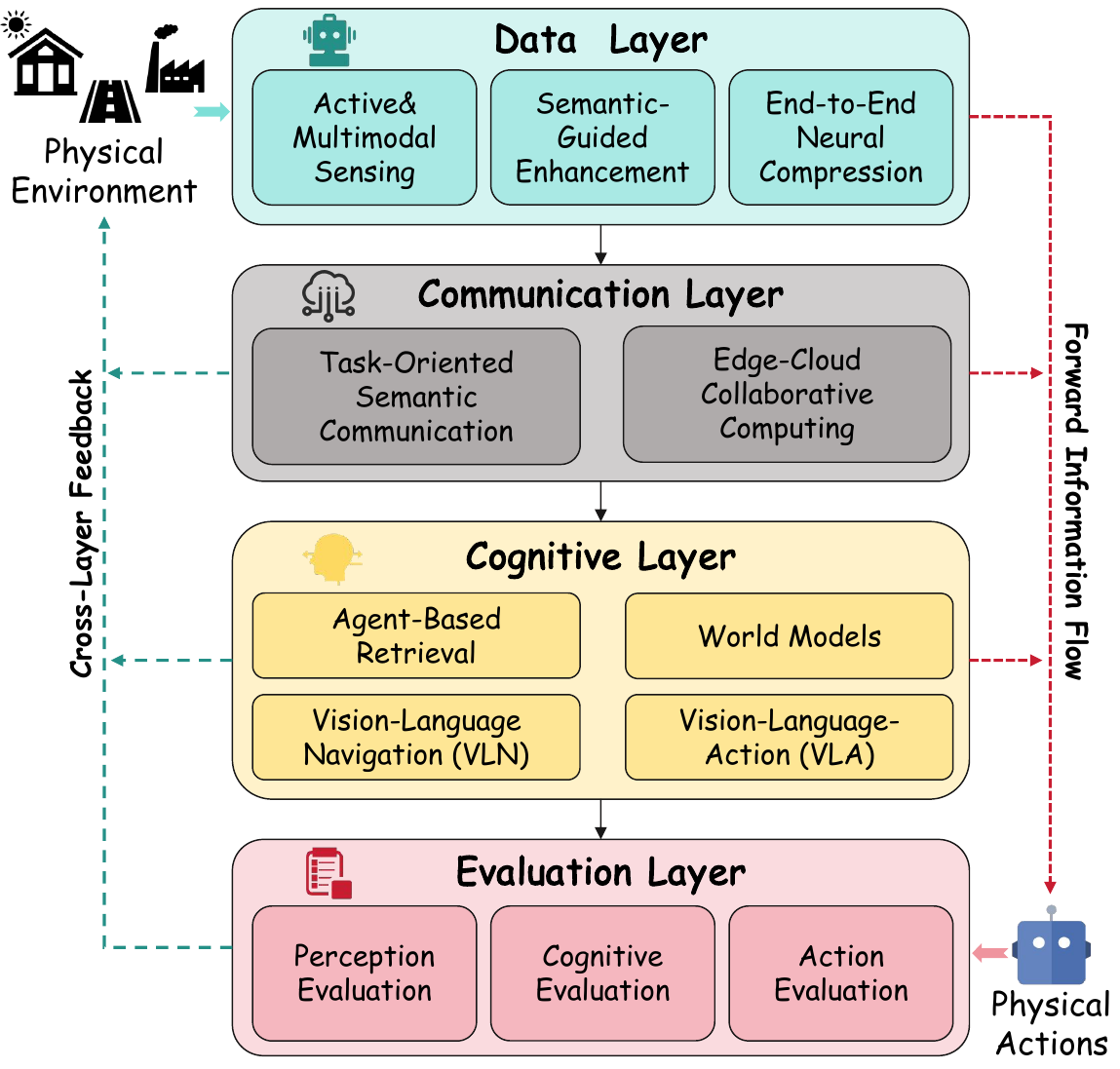}
\caption{The proposed four-layer Embodied Multimedia architecture.}
\label{fig:arch}
\vspace{-15pt}
\end{figure}

The proposed four-layer architecture is illustrated in Fig.~\ref{fig:arch}. The layers reflect the flow of information from raw sensor data to physical execution and evaluative feedback, with cross-layer interactions enabling continuous adaptation.

The \textit{Data Layer} provides the perceptual foundation. It encompasses active multimodal sensing with adaptive viewpoint selection, semantic-guided enhancement that restores task-relevant features in degraded inputs, and end-to-end neural compression that preserves semantically important information at reduced bitrates.
The \textit{Communication Layer} connects edge perception devices to cloud-side cognitive systems. Departing from bit-level transmission, it adopts task-oriented semantic communication paradigms that encode and relay task-relevant information efficiently under constrained bandwidth. Edge-cloud collaborative computing \cite{liu2026edge} distributes the inference workload hierarchically to satisfy real-time control requirements.
The \textit{Cognitive Layer} serves as the decision-making core. It integrates agent-based retrieval for dynamic knowledge access, world models for environmental prediction and counterfactual planning, Vision-Language Navigation (VLN) for instruction-guided spatial reasoning, and Vision-Language-Action (VLA) models that map multimodal inputs directly to physical control signals.
The \textit{Evaluation Layer} establishes assessment criteria aligned with embodied task objectives. It covers machine preference-driven perception evaluation, multi-dimensional cognitive benchmarks encompassing spatial reasoning and safety, and action evaluation through large-scale navigation and manipulation benchmarks.


\section{Data Layer}
\label{sec:data}

\subsection{Active and Multimodal Sensing}

Passive data collection from fixed sensors is insufficient for agents operating in dynamic, partially observable environments. Active perception enables agents to compute maximally informative viewpoints, actively resolving occlusions and directing attention toward semantically relevant regions. Simulation platforms such as Habitat~\cite{savva2019habitat} provide environments for training and evaluating active sensing policies at scale before physical deployment. Event cameras, which respond asynchronously to local intensity changes rather than producing synchronized full frames, reduce data redundancy and provide high temporal resolution in fast-motion scenarios, complementing conventional RGB sensing.

Tactile sensing extends the perceptual modality space beyond vision. Piezoresistive, capacitive, and vision-based tactile sensors each offer distinct tradeoffs among resolution, cost, and physical robustness. Tactile feedback provides contact force distribution, surface texture, and object geometry that are inaccessible to vision under occlusion, such as grasping pressure and material compliance during in-hand manipulation. Fusing tactile signals with visual and inertial inputs enables substantially more reliable manipulation of deformable and geometrically novel objects. Multimodal sensing across vision, touch, and proprioception thus forms the perceptual cornerstone of the Data Layer.

Embodied Multimedia also addresses training data scarcity through physics-based synthetic data generation. Digital human motion generation methods synthesize diverse, semantically consistent movement sequences, providing large-scale training data for agents that must coordinate with people. Physics simulators produce interaction scenarios with configurable material properties and contact dynamics, enabling the generation of data distributions that would be prohibitively expensive to collect in the real world.

\subsection{Semantic-Guided Enhancement}

Enhancement in Embodied Multimedia departs from uniform pixel-level optimization and instead applies selective restoration guided by semantic understanding. By exploiting priors encoded in large pretrained vision-language models such as CLIP~\cite{radford2021learning}, enhancement networks identify and prioritize task-relevant regions rather than processing the image globally. Degradation-aware controllers trained through vision-language alignment recognize the type of damage present and execute targeted restoration. Text-to-image diffusion models contribute semantic embeddings as generation constraints during super-resolution, ensuring that restored textures conform to physically plausible semantic relationships, thereby providing reliable inputs for downstream perception and manipulation.

\subsection{End-to-End Neural Compression}

Neural compression architectures based on variational autoencoders have surpassed traditional codecs in rate-distortion performance on standard benchmarks. Task-driven variants incorporate semantic loss terms to retain features critical for object detection and segmentation even at aggressive compression ratios. Entropy models based on Transformer attention mechanisms further improve probability estimation, reducing residual redundancy. For video, conditional coding architectures such as DCVC~\cite{li2021dcvc} exploit temporal feature-space context rather than explicit optical flow, reducing artifacts while improving computational efficiency. Mobile-optimized variants achieve real-time 1080p video decoding on resource-constrained hardware, directly enabling Embodied Multimedia pipelines on mobile robot platforms.

\section{Communication Layer}
\label{sec:comm}

\subsection{Semantic Communication}

Semantic communication~\cite{xie2021deepsc} shifts the fundamental transmission objective from reliable reconstruction of every transmitted bit to the preservation of task-relevant meaning. End-to-end trained joint source-channel coding systems map semantic representations directly to channel-adapted signals, eliminating the explicit separation between source and channel coding characteristic of traditional architectures. For text, attention-based semantic encoders extract task-relevant linguistic features while minimizing semantic error under noisy channel conditions. For image and multimodal data, codebook-based vector quantization achieves compact semantic representations that remain reliable under bandwidth constraints. In embodied scenarios, goal-oriented communication architectures further constrain encoding to information required for downstream task decisions, substantially reducing transmission overhead while maintaining control quality~\cite{zhang2022toward6g}.

The convergence of semantic communication with B5G network design offers a transformative opportunity. Semantic-aware network slicing allocates bandwidth according to task criticality rather than raw data volume, prioritizing safety-critical control signals over lower-priority sensor streams. Hybrid retransmission protocols encode incremental semantic information for error correction, maintaining communication reliability without reverting to full bit-level retransmission, which are particularly valuable for embodied systems that must operate reliably under fluctuating network conditions.

\subsection{Edge-Cloud Collaboration}

The latency constraints of closed-loop robotic control preclude sole reliance on remote cloud inference, while edge devices lack the computational budget for full-scale foundation models. End-Edge-Cloud Computing (EECC) \cite{liu2026edge} hierarchies distribute computation across three tiers: terminal devices handle lightweight feature extraction and real-time low-level control; edge nodes run locally deployed medium-scale models for immediate reactive response; cloud servers support complex planning, global model training, and knowledge distillation back to edge-deployed models. The hierarchical distribution reduces control latency while preserving access to high-capacity models for long-horizon reasoning. Federated learning mechanisms enable privacy-preserving collaborative model updates across distributed embodied platforms. 

\section{Cognitive Layer}
\label{sec:cog}

\subsection{Agent-Based Retrieval}

Static retrieval systems that match queries against fixed feature indexes are insufficient for embodied tasks requiring multi-step reasoning over knowledge-intensive instructions in open environments. Agent-based retrieval establishes a dynamic closed loop of four phases: task analysis, retrieval execution, result optimization, and strategy iteration. In the task analysis phase, the agent decomposes complex or ambiguous queries into structured sub-queries, reducing semantic ambiguity. The execution phase combines sparse term-based methods for exact lexical matching with dense semantic methods for conceptual similarity, exploiting the complementary strengths of both retriever types. Retrieved content is reranked to remove noise before being provided to the reasoning module. When initial results are insufficient, the agent reformulates queries based on generation feedback and iterates until the retrieved knowledge adequately supports the current decision.

\subsection{World Models}

World models enable agents to simulate future environmental states, supporting planning and risk evaluation without requiring physical trial and error. Generative world models produce future visual observations conditioned on current state and action sequences~\cite{bar2025nav}, allowing agents to evaluate candidate action plans in a simulated world before committing to physical execution. Representation-prediction models such as V-JEPA~2 operate in latent feature space rather than pixel space, predicting future state representations without the computational overhead of high-fidelity image synthesis. This abstraction improves robustness to visual distractors and reduces computation. DreamerV3~\cite{hafner2025dreamer} demonstrates that a single latent dynamics model trained across diverse task domains generalizes to new control problems without domain-specific tuning, which is a notable step toward world models for general embodied intelligence. In autonomous driving, multi-view spatiotemporal world models extend prediction to complex traffic scenarios for safety-aware trajectory planning.

\subsection{Vision-Language Navigation}

VLN requires embodied agents to execute navigation sequences through visual environments guided solely by natural language instructions, capturing the core challenge of grounding abstract linguistic directives in perceptual experience. Environment model-based approaches maintain explicit spatial representations, including topological graphs and volumetric occupancy maps, to support long-range memory and systematic planning. Reasoning-based approaches forego explicit map construction and instead leverage large vision-language models to decompose instructions into actionable sub-goals and recover from errors through language-guided backtracking. Recent methods incorporate world models into the navigation loop, enabling agents to predict future visual observations before committing to movement, improving performance in previously unseen environments.

\subsection{Vision-Language-Action Models}

VLA models constitute the final stage that translates multimodal perceptual inputs into physical control outputs. The dominant action tokenization paradigm, demonstrated at scale by RT-1~\cite{brohan2022rt1}, encodes robotic actions as discrete tokens within a vocabulary shared with language and visual tokens. This representation allows Transformer architectures pretrained on large vision-language corpora to be adapted for manipulation through fine-tuning, transferring rich semantic knowledge to physical control. RT-2~\cite{zitkovich2023rt2} further demonstrated that internet-scale visual-language pretraining improves zero-shot generalization to novel objects and instruction types. The $\pi_0$ model~\cite{black2024pi0} replaces discrete token prediction with flow matching, producing continuous high-frequency action trajectories directly compatible with dexterous manipulation. FAST~\cite{pertsch2025fast} applies frequency-domain tokenization based on the discrete cosine transform, compactly representing complex motion profiles and improving training efficiency on fine-grained tasks.

Beyond direct action output, complementary paradigms generate intermediate representations. Affordance and trajectory prediction methods provide spatial constraints for low-level controllers, bridging abstract semantics with precise 3D interaction targets. Code generation approaches translate natural language instructions into executable programs, enabling compositional and interpretable task plans. Closed-loop reasoning frameworks maintain linguistic summaries of the execution state, allowing policies to adapt their behavior in response to environmental feedback during long-horizon task execution.

\section{Evaluation Layer}
\label{sec:eval}

\subsection{Perception Evaluation}

The Machine Preference framework~\cite{li2025machine} redefines image quality by the performance consistency of downstream machine vision tasks, such as object detection and semantic segmentation, rather than by human perceptual ratings. A large-scale Machine Preference Database containing over 2.25 million annotated samples demonstrates that conventional perceptual metrics correlate poorly with machine task performance. Region-aware quality assessment further identifies spatial degradation patterns that affect machine perception disproportionately, providing precise diagnostic information for the design of enhancement and compression systems oriented toward machine utility rather than human viewing.

\subsection{Cognitive Evaluation}

Cognitive evaluation spans basic scene understanding, interaction planning, and safety assessment. Embodied question answering established the paradigm of requiring agents to navigate and reason about their environment to answer language questions. Subsequent benchmarks extended this to 3D point cloud scenes, position-aware reasoning, and physical common-sense understanding of object dynamics and forces~\cite{chow2025physbench}. Safety-oriented benchmarks evaluate agent under hazardous or adversarial conditions~\cite{lu2025isbench,liu2025agentsafe}, testing risk identification, instruction compliance, and  mitigation execution.

\subsection{Action Evaluation}

Navigation benchmarks beginning with R2R have progressively increased task difficulty by adding longer trajectories, multilingual instructions, and object referencing in previously unseen environments. Manipulation benchmarks including RLBench~\cite{james2020rlbench}, ManiSkill~\cite{mu2021maniskill}, and LIBERO~\cite{liu2023libero} provide scalable task suites spanning grasping, assembly, and sequential manipulation across varying task horizons and prior knowledge levels. Open X-Embodiment~\cite{oneill2024oxe} integrates demonstrations from 22 distinct robot types across 21 institutions, enabling evaluation of cross-platform policy generalization. 

\section{Challenges and Frontier Directions}
\label{sec:challenges}

\subsection{Technical Challenges}

Despite significant recent progress, several fundamental challenges impede the deployment of Embodied Multimedia systems across the four architectural layers.

At the Data Layer, multimodal spatiotemporal alignment across heterogeneous sensors remains unsolved at the millisecond timescales required for physical interaction. Cameras, LiDAR, tactile arrays, and inertial units differ substantially in sampling rate and signal structure, complicating joint representation learning. High-quality embodied interaction data capturing causal relationships, failure modes, and rare long-tail events is also scarce and expensive to collect, limiting generalization to out-of-distribution scenarios.

At the Communication Layer, network unreliability introduces vulnerabilities for edge-cloud offloading architectures. Bandwidth fluctuations and packet loss degrade control continuity in dynamic environments, and the emergence of Networked AI~\cite{song2022networking}, in which multiple agents share computational resources and learned representations across network nodes, further demands co-design of communication protocols and AI inference pipelines. The Networking System of AI \cite{song2022networking} requires semantic-aware routing that adapts to the collective task state of a robot fleet, not merely individual link conditions.

At the Cognitive Layer, the sim-to-real gap persists as a primary deployment barrier. Physics simulators  provide imperfect representations of contact mechanics and material compliance, and policies that succeed in simulation regularly fail due to distribution shifts in real environments. Concurrently, foundation-model-based agent systems, including tool-use architectures that equip language models with executable skill libraries, are beginning to address open-world task generalization. Continual and online learning mechanisms that allow agents to refine their skill repertoire from real deployment experience, without catastrophic forgetting, remain an open problem central to the Cognitive Layer.

At the Evaluation Layer, hardware constraints on mobile platforms create a persistent tension between model capacity and real-time latency. VLA and world models demand substantial computation, while mobile robots operate under strict power and thermal budgets. Model compression and parameter-efficient fine-tuning are necessary to enable sustained on-device inference, and standardized cross-platform evaluation protocols that measure performance under realistic hardware constraints remain underdeveloped.

\subsection{Frontier Applications}

Table~\ref{tab:frontier} summarizes five frontier directions where Embodied Multimedia is positioned to serve a foundational role.

\begin{table}[t]
\centering
\caption{Frontier Application Directions for Embodied Multimedia}
\label{tab:frontier}
\setlength{\tabcolsep}{3pt}
\renewcommand{\arraystretch}{1.25}
\begin{tabular}{p{0.2cm}p{3.5cm}p{4.4cm}}
\toprule
 & \textbf{Key Challenge} & \textbf{Role of EMM} \\
\midrule
(a) & Semantic delivery, low latency   & Goal-oriented coding, B5G networks \\
(b) & Physics-grounded reasoning       & Physics-aware world models \\
(c) & Multimodal evidence fusion        & Cross-modal anomaly reasoning \\
(d) & Physical-virtual integration      & Digital twins, mixed-reality pipelines \\
(e) & Aesthetic-physical alignment      & Generative models with VLA control \\
\bottomrule
\end{tabular}
\vspace{-18pt}
\end{table}

\textit{(a) Multimedia Communication.}
Semantic communication frameworks aligned with B5G communication~\cite{zhang2022toward6g} transmit task-relevant meaning rather than raw bitstreams, matching the low-latency, high-reliability demands of embodied control. Beyond single-agent systems, the Networked AI paradigm envisions fleets of embodied agents sharing compressed semantic representations and learned skills across a distributed network, requiring joint co-design of communication protocols and AI inference pipelines. Network-aware adaptive encoding that adjusts compression and transmission priority based on collective task state represents a central open challenge. 

\textit{(b) Physical Intelligence.}
Physical intelligence requires reasoning about object mass, friction, deformation, and contact forces in a way that current VLA models do not yet support \cite{qian2026survey}. Agent AI systems that equip foundation models with callable skill libraries offer a promising direction, allowing agents to invoke specialized physical manipulation routines on demand \cite{liu2026understanding}. EMM must supply the supporting data and evaluation infrastructure: interaction datasets annotated with material parameters and contact forces, and benchmarks such as PhysBench~\cite{chow2025physbench} that measure physical reasoning fidelity.

\textit{(c) Embodied Anomaly Perception.}
Industrial inspection and safety surveillance demand embodied agents that detect anomalies imperceptible to passive cameras, including subsurface defects revealed through thermal imaging or mechanical faults sensed through contact profiles \cite{liu2025crcl}. Active probing enables agents to direct sensors toward suspicious regions and fuse heterogeneous evidence over time. Online Evolutive Learning (OEL) \cite{qian2026coordinated} mechanisms that continuously update anomaly models from deployment feedback, without full retraining, are essential for practical deployment. Principled multimodal anomaly benchmarks for physical inspection, informed by emerging safety evaluation frameworks~\cite{lu2025isbench,liu2025agentsafe}, remain scarce and represent an important open direction.

\textit{(d) Metaverse and Interactive Multimedia.}
Persistent virtual environments populated by both human users and autonomous agents require multimedia pipelines that bridge physical and digital interaction \cite{wang2025immersive}. Augmented and mixed reality systems~\cite{billinghurst2015ar} overlay perceptual and actuation capabilities onto physical space, while digital twins synchronize virtual replicas with physical state in real time. Embodied agents in these settings must coordinate across both domains simultaneously. Latency-bounded synchronization, physically grounded avatar animation, and the integration of emerging input modalities such as neural interfaces and full-body haptics extend the scope of both the Data and Communication Layers.

\textit{(e) Art Creation and Generative Physical Expression.}
Creative embodied systems that paint, sculpt, or perform music must integrate aesthetic reasoning with precise manipulation. Generative models have demonstrated strong creative capacity in the digital domain; extending this to physical media requires embodied perception of material properties and tool-surface interaction. Online learning from human feedback during co-creative sessions offers a route toward agents that adapt their expressive style in real time. 

\section{Conclusion}
\label{sec:conclusion}

This paper formally introduces Embodied Multimedia as a cross-disciplinary research paradigm addressing the structural mismatch between human-centric multimedia infrastructure and the perception-decision-action demands of embodied intelligent agents. The proposed four-layer architecture, spanning Data, Communication, Cognitive, and Evaluation layers, provides an organizing conceptual framework for EMM research and application. The survey of enabling technologies across each layer illustrates both the breadth of relevant research contributions and the depth of challenges that remain. As embodied agents advance toward more capable, general, and safe physical interaction with the world, EMM will serve as a foundational paradigm for the data, communication, and evaluation infrastructure that enables interactive perception, decision-making, and action in the physical  and virtual word.

\vspace{-5pt}
\bibliographystyle{IEEEtran}
\bibliography{refs}

\end{document}